\documentclass[10pt,notitlepage,aps,pra,superscriptaddress,floatfix,nofootinbib,twocolumn]{revtex4-2}

\usepackage{amsmath,amsfonts,amsthm,amssymb}
\usepackage[utf8]{inputenc}
\usepackage[T1]{fontenc}
\usepackage{graphicx}
\usepackage{xspace}
\usepackage[colorlinks=true,citecolor=blue,urlcolor=magenta,linkcolor=cyan]{hyperref}
\usepackage[capitalize]{cleveref}
\usepackage{bbm}
\usepackage{bm}
\usepackage{xcolor}
\usepackage{xfrac}
\usepackage{float}

\crefname{section}{Sec.}{Secs.}
\crefname{equation}{Eq.}{Eqs.}
\crefname{figure}{Fig.}{Figs.}
\crefname{appendix}{Appendix}{Appendices}

\newcommand{\one}{\mathbbm{1}} 
\newcommand{\id}{\text{id}} 

\newcommand{\cA}{\mathcal{A}}
\newcommand{\cE}{\mathcal{E}}
\newcommand{\cI}{\mathcal{I}}

\let\epsilon\varepsilon
\newcommand{\T}{\text{T}}
\newcommand{\va}{\vec{a}}

\newcommand{\ee}{\mathrm{e}}
\newcommand{\ii}{\mathrm{i}}
\newcommand{\Tr}[1]{\operatorname{tr}\left[ #1 \right]}
\newcommand{\ket}[1]{\left \vert #1 \right \rangle}
\newcommand{\bra}[1]{\left \langle #1 \right \vert}

\newcommand{\ketbra}[2]{\left \vert #1 \right \rangle \left \langle #2 \right \vert}
\newcommand{\purestate}[1]{\left \vert #1 \right \rangle \left \langle #1 \right \vert}
\renewcommand{\vec}[1]{\boldsymbol{#1}}

\newcommand{\DC}{\operatorname{DC}}

\begin{document}
\title{Entanglement-breaking channels are a quantum memory resource}

\author{Lucas B. Vieira}
\email{lucas.vieira@tu-darmstadt.de}
\affiliation{Department of Computer Science, Technical University of Darmstadt, Darmstadt, Germany}
\affiliation{Institute for Quantum Optics and Quantum Information (IQOQI), Austrian Academy of Sciences,\\ Boltzmanngasse 3, 1090 Vienna, Austria}
\affiliation{Faculty of Physics, University of Vienna, Boltzmanngasse 5, 1090 Vienna, Austria}

\author{Huan-Yu Ku}
\email{huan.yu@ntnu.edu.tw}
\affiliation{Department of Physics, National Taiwan Normal University, Taipei 11677, Taiwan}

\author{Costantino Budroni}
\email{costantino.budroni@unipi.it}
\affiliation{Department of Physics ``E. Fermi'' University of Pisa, Largo B. Pontecorvo 3, 56127 Pisa, Italy}
\affiliation{Faculty of Physics, University of Vienna, Boltzmanngasse 5, 1090 Vienna, Austria}
\affiliation{Institute for Quantum Optics and Quantum Information (IQOQI), Austrian Academy of Sciences,\\ Boltzmanngasse 3, 1090 Vienna, Austria}

\date{\today}

\begin{abstract}
Entanglement-breaking channels (equivalently, measure-and-prepare channels) are quantum operations noted for their ability to destroy multipartite spatial quantum correlations. Inspired by this property, they have also been widely employed in defining notions of ``classical memory'', under the assumption that such channels effectively act as a classical resource. In this paper, we show this assumption is false. By means of multi-time correlations, we conclusively show that entanglement-breaking channels are still a quantum resource: a qudit going through an entanglement-breaking channel can generate genuinely nonclassical temporal correlations, i.e., that cannot be simulated by a classical system of same dimension. Our results imply that entanglement-breaking channels cannot generally be employed to characterize classical memory effects without additional assumptions.
\end{abstract}

\maketitle

\section{Introduction}
Classical memory, in the sense of storage of information, is a familiar notion to us from daily life. With the advent of quantum information and quantum computation, several analogous notions of \emph{quantum memory} have now been introduced to describe quantum resources offering advantages over their classical counterparts. These advantages have been attributed to disparate quantum phenomena in a variety of scenarios~\cite{jozsa2003role,vedral2010_elusive,dakic2014_discord,pirandola2014_discord,mansfield2018_contextuality,howard2014,saha2023_incompatibility}, making the essential features for a quantum memory the subject of ongoing debate. Nevertheless, even if we lack definitive criteria for the ``quantumness'' of a memory, its ``classicality'' should unambiguously indicate a total absence of any quantum advantages. Thus, a complete understanding of \emph{classical} memory effects becomes essential to any practical notion of \emph{nonclassical} memory effects resulting from a quantum memory.

Quantum memories feature prominently in the study of quantum networks~\cite{rosset2018_resourcetheory,Simnacher2019PRA,Graffitti2020_PRL,giarmatzi2021_witnessingquantum} and multi-time quantum processes~\cite{milz2020_whenis,berk2021_resourcetheoriesof,taranto2023_hierarchy}. The notion of classical and quantum memory also plays a role in open quantum systems~\cite{morris2022_quantifying,shrikant2023_quantum}, where the environment acts as the memory resource leading to temporally-correlated noise, and in notions of classical and quantum communication capacities~\cite{frenkel2015_storage}.
Clearly, a precise characterization of classical and quantum memories is of great theoretical and practical interest. To this end, the formalism of quantum supermaps~\cite{chiribella2008_supermaps,pollock2018_processes,pollock2018_nonmarkovian} (also known as process tensors or quantum combs) has been extensively applied to the investigation of quantum processes~\cite{milz2020_whenis,berk2021_resourcetheoriesof,taranto2023_hierarchy,shrikant2023_quantum}, where arbitrary protocols can be reinterpreted as multipartite quantum states. For this reason, several notions of ``classical memory'' have been proposed in analogy to spatial correlations, often in conflict with one another~\cite{taranto2023_hierarchy}, but nevertheless all based on interpreting separability of a supermap as a signature of the classicality of its memory~\cite{rosset2018_resourcetheory,berk2021_resourcetheoriesof,giarmatzi2021_witnessingquantum,taranto2023_hierarchy,backer2024_localdisclosure}.

Building on this, the typical approach to render the memory in a quantum supermap ``classical'' employs entanglement-breaking (EB) channels, known to be exactly equivalent to measure-and-prepare operations~\cite{horodecki2003_ebchannels}. As the input quantum state is destroyed by the measurement and replaced by a classical label (albeit encoded in the prepared output quantum state), one may be inclined to claim that such channels are essentially equivalent to classical channels~\cite{rosset2018_resourcetheory,Simnacher2019PRA,yuan2021_benchmarking,ku2022_quantifying,Abiuso_2024,berk2021_resourcetheoriesof,santos2024_quantifying}, i.e., that they can only transmit classical information between their inputs and outputs, therefore being restricted to classical correlations. EB channels have been used in this manner, e.g., to develop resource theories of quantum memories~\cite{rosset2018_resourcetheory,berk2021_resourcetheoriesof}, the benchmarking of quantum devices~\cite{yuan2021_benchmarking}, characterizing memory effects in quantum processes~\cite{giarmatzi2021_witnessingquantum,milz2021_genuine,berk2021_resourcetheoriesof,roy2024_semidevice,santos2024_quantifying} and bosonic Gaussian channels~\cite{ivan20213_ncbreaking,Abiuso_2024}, and in establishing a hierarchy between competing notions of quantum processes with classical memory~\cite{taranto2023_hierarchy}.

The issue with these approaches is that they still operate within the quantum formalism, obscuring the distinction between quantum and classical memory effects, and thereby preventing a proper characterization of these memory resources. Instead, the distinction between classical and quantum memory effects should emerge from explicit constructions within each theory. Thus, a more rigorous approach consists in establishing bounds on information processing tasks achievable with classical memory alone. A violation of these bounds then provides a certificate of nonclassicality, often also providing a quantifiable metric for the \emph{quantumness} of the underlying memory resource. Crucially, this requires not only the most general possible notion of ``classical memory'', but also an appropriate choice of a task where classical and quantum memories can be compared on equal terms.

In this paper, we show that interpreting EB channels generically as classical memory resources can be  misguided. We address the following question: is the memory of a dit (a classical system with $d$ states) equivalent to that of a \emph{qudit} ($d$-level quantum system) passing through an EB channel? We answer it in the negative, by presenting explicit scenarios where a qudit passing through an EB channel outperforms a dit of classical memory in the task of generating correlations in time. We conclude that, contrary to prevailing interpretations and applications, EB channels cannot generically be used to define or characterize classical memory effects in multi-time scenarios.

\begin{figure}[t]\centering
	\includegraphics[width=0.9\linewidth]{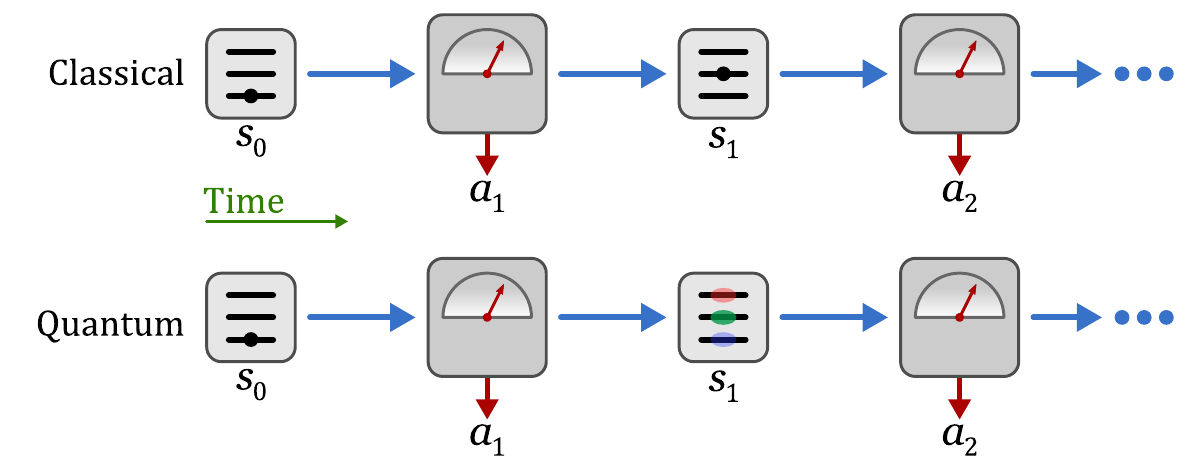}
	\caption{Temporal correlations can be investigated by a classical or quantum system with $d$ distinguishable internal states, in the initial state $s_0$, which is repeatedly measured by the same instrument at multiple times. The measurement of state $s_t$ produces the outcome $a_{t+1}$ and changes the state to $s_{t+1}$.}
	\label{fig:idea}
\end{figure}

\section{Preliminary notions}
\begin{figure}[t]\centering
	\includegraphics[width=1.0\linewidth]{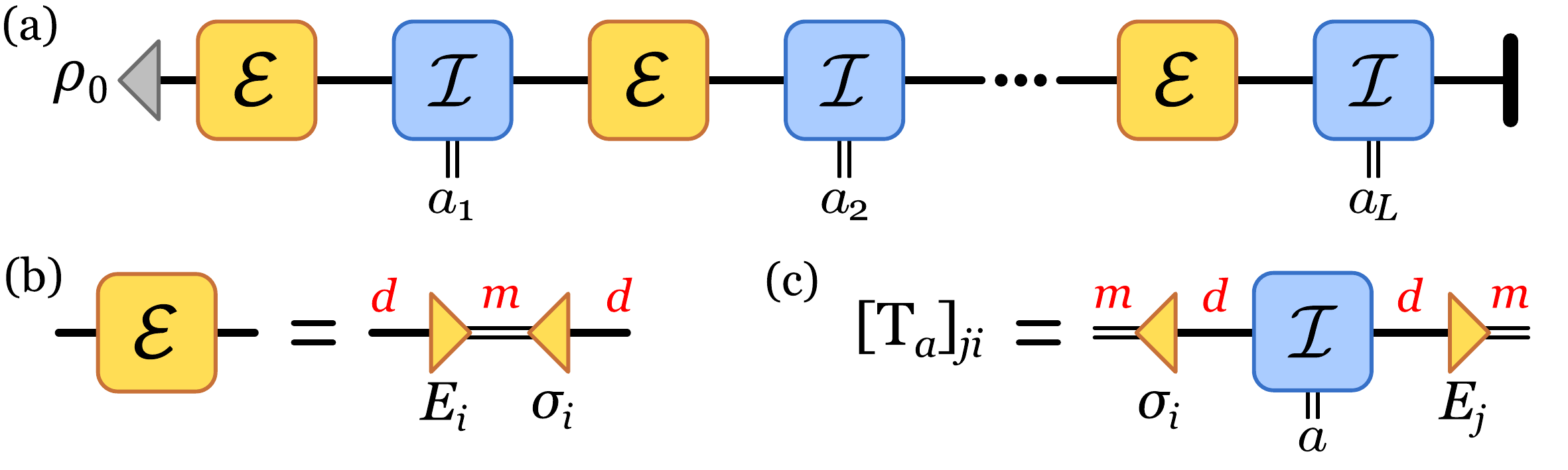}
	\caption{(a) The sequential measurement protocol considered in this work. An isolated quantum system in an initial state $\rho_0$ is repeatedly passed through an entanglement-breaking channel $\cE$ before being measured by a quantum instrument $\cI$, obtaining a sequence of outcomes $\va = a_1 a_2 \dots a_L$. (b) The entanglement-breaking channel $\cE$ understood as a measure-and-prepare operation on a $d$-dimensional quantum system, with a $m$-dimensional intermediate classical space. (c) The $m \times m$ transition matrix of the effective classical model.}
	\label{fig:protocol}
\end{figure}

A channel $\cE_A$ is said to be \emph{entanglement-breaking} if $(\cE_A \otimes \id_B)(\rho_{AB})$ is separable for any bipartite state $\rho_{AB}$. A well-known result establishes that any EB channel $\cE$ can be written as a measure-and-prepare channel in the Holevo form~\cite{holevo1998_qcodingthms,horodecki2003_ebchannels}
\begin{equation}\label{eq:EBCTr}
	\cE(\rho) = \sum_{i=1}^{m} \Tr{\rho E_i} \sigma_i,
\end{equation}
for $\{E_i\}_{i=1}^m$ a positive operator-valued measure (POVM), i.e., $E_i \ge 0$ and $\sum_{i=1}^m E_i = \one$, and $\sigma_i$ quantum states. Here, $m$ can be interpreted as the number of labels---in principle, arbitrarily large---required in the channel's intermediate classical memory in order to describe its operation~\cite{,horodecki2003_ebchannels,rosset2018_resourcetheory}. These channels can be understood in terms of a quantum-classical-quantum operation:
(1) The input quantum state $\rho$ is first measured, producing a classical outcome label $i$, out of a total $m$, with probability $\Tr{\rho E_i}$,
(2) Conditioned on $i$, the corresponding quantum state $\sigma_i$ is passed forward as output; see Fig.~\hyperref[fig:protocol]{\ref{fig:protocol}b}. EB channels can be fully characterized~\cite{horodecki2003_ebchannels} by the fact their Choi matrix~\cite{jamiolkowski1972,choi1975}, $C_\cE := (\cE \otimes \id)(\purestate{\Phi})$ with $\ket{\Phi}$ a maximally entangled state, is separable and with rank $r \ge d$, which also implies any such channel can be rewritten with at most $m = d^2$ operators $E_i$ and $\sigma_i$.

To precisely distinguish between classical and quantum memories, we investigate their capabilities when generating temporal correlations in the simplest scenario, i.e., that of a single, finite-dimensional system subject to the same dichotomous measurement at multiple times, as shown in \Cref{fig:idea}. This scenario can be described within the formalism of finite-state machines~\cite{budroni2019_memorycost,budroni2021_tickingclocks,vieira2022_temporal}, recently introduced to the study of temporal correlations as it allows an exact characterization of classical, quantum, and generalized probability theories~\cite{budroni2019_memorycost}.

In the classical case, the most general finite-state machine description is given by an initial stochastic vector $\pi$ over its $d$ distinct states (i.e., a dit of memory), and a pair of (column) sub-stochastic transition matrices $T = (T_0, T_1)$ describing its dynamics upon each outcome $a \in \{0,1\}$, with $T_0 + T_1$ column stochastic. Importantly, this description allows classical measurements to be invasive, going beyond typical definitions of classicality rooted on unnecessarily macroscopic notions of realism, as in Leggett-Garg inequalities~\cite{leggett1985_quantum,zukowski2014_temporal,brierley2015_nonclassicality,vitagliano2023_leggettgarg} and in the application of the Kolmogorov extension theorem to stochastic processes~\cite{milz2020_kolmogorovextension}. The probability of a length-$L$ sequence of outcomes $\va = a_1 a_2 \ldots a_L$ is then given by
\begin{equation}\label{eq:pC}
	p(\va | T, d) = \eta T_{a_L} \cdots T_{a_2} T_{a_1} \pi,
\end{equation}
with $\eta = (1, \dots, 1)$. The quantum case is defined analogously, with a single $d$-dimensional quantum system (a qudit), initially at state $\rho_0$, being repeatedly measured by a quantum instrument $\cI = (\cI_0, \cI_1)$, with effects $\cI_a$ being completely positive (CP) trace non-increasing maps such that $\cI_0 + \cI_1$ is completely positive trace-preserving (CPTP). The probability of a sequence $\va$ is given by
\begin{equation}\label{eq:pQ}
	p(\va | \cI, d) = \Tr{\cI_{a_L} \circ \cdots \circ \cI_{a_1}(\rho_0)}.
\end{equation}
In both classical and quantum cases, the system's state is the sole memory resource available to generate the probability distributions, and any dynamics is entirely governed by the effects associated with a measurement outcome $a_t$. In this way, classical and quantum memories are placed on equal terms as a memory resource.

A key observation is that memory is the main resource for realizing the distributions in \cref{eq:pC,eq:pQ}, as any temporal correlation can be generated by a classical or quantum system with enough memory, i.e., enough internal states~\cite{fritz2010_temporalCHSH,clemente2016,hoffmann2018_temporalqubit, spee2020_simulating}. Consequently, the system's dimension establishes fundamental limitations on the set of achievable temporal correlations~\cite{hoffmann2018_temporalqubit,budroni2019_memorycost,spee2020_simulating, budroni2021_tickingclocks,vieira2022_temporal}, providing an unambiguous notion for distinguishing between classical and quantum memories. In other words, bounding the amount of memory is absolutely essential for establishing any meaningful notion of classicality or nonclassicality of temporal correlations. Note that we do not allow external storage of information (e.g., by conditioning on past outcomes, the use of shared randomness between time steps, or invoking an external reference clock), ensuring that the total amount of memory can be precisely accounted for by the number of orthogonal states $d$ of the memory system.

One way to distinguish between classical and quantum memory effects is by establishing upper bounds on the task of \emph{sequence generation}, i.e., the maximum probability for individual sequences to be produced within classical and quantum theory~\cite{budroni2019_memorycost,budroni2021_tickingclocks,vieira2022_temporal}. The notion of \emph{deterministic complexity} of a sequence $\va$, denoted by $\DC(\va)$, was introduced in Ref.~\cite{vieira2022_temporal} as the minimum dimension $d$ for a system (classical or quantum) to be capable of generating $\va$ with probability one. Thus, if $d < \DC(\va)$, then no classical or quantum system can produce the sequence $\va$ deterministically, implying both classical and quantum theory must obey nontrivial upper bounds for the maximum probability. Since $d = 2$ is the smallest nontrivial amount of memory, these observations also highlight the importance of studying temporal correlations---and thus, memory effects---in scenarios with three or more time steps, as every sequence of length two can be produced deterministically with either a bit or qubit.

As the underlying memory resource can be explicitly confined within each theory, such nontrivial upper bounds provide a sharp demarcation between classical and quantum memories and their capabilities, but they are generally difficult to obtain precisely~\cite{vieira2022_temporal,weilenmann2023_optimisation,vieira2023_witnessing}. In Ref.~\cite{vieira2022_temporal}, a universal upper bound of $1/\ee$ was conjectured for all sub-deterministic classical memory scenarios, whereas no nontrivial universal quantum bound seems to exist.

\section{The sequential measurement protocol}
We investigate the nonclassicality of entanglement-breaking channels by inserting a fixed EB channel $\cE$, as in \Cref{eq:EBCTr}, before each quantum measurement in \Cref{eq:pQ} (depicted in Fig.~\hyperref[fig:protocol]{\ref{fig:protocol}a}), obtaining
\begin{align}\label{eq:probQ}
\begin{split}
	&p(\va|\cI, \cE, d) = \Tr{\cI_{a_L} \circ \cE \circ \cdots \circ \cI_{a_2} \circ \cE \circ \cI_{a_1} \circ \cE(\rho_0)} \\
	&= \sum_{i_1, \ldots, i_{L} = 1}^{m} \Tr{ \cI_{a_{L}}(\sigma_{i_{L}}) \one } \cdots \Tr{ \cI_{a_1}(\sigma_{i_1}) E_{i_2} } \Tr{\rho_0 E_{i_1}}  \\
	&= \sum_{i_1, \ldots, i_{L+1} = 1}^{m} \eta_{i_{L+1}} [T_{a_L}]_{i_{L+1} i_{L}} \cdots [T_{a_1}]_{i_2 i_1}  \pi_{i_1},
\end{split}
\end{align}
where we use $[T_a]_{ji} := \Tr{ \cI_a(\sigma_i) E_{j} }$, $\pi_{i} := \Tr{\rho_0 E_{i}}$, and $\eta = (1, \dots, 1)$ as in \Cref{eq:pC}. Note that, since the $E_i$ are POVM elements, each trace term corresponds to a transition probability. We have thus collected all such probabilities in the matrices $T_a$ such that
\begin{equation}\label{eq:ItoT}
	\sum_{a,j} [T_a]_{ji} = \Tr{ \textstyle{\sum_{a}} \cI_{a}(\sigma_{i}) \cdot \textstyle{\sum_{j}} E_{j} } = \Tr{ \hat{\sigma}_i \one } = 1,
\end{equation}
where we use the fact $E_i$ form a POVM and $\sum_{a} \cI_{a}$ is CPTP, thus $\hat{\sigma}_i = \sum_{a} \cI_{a}(\sigma_i)$ has unit trace. Furthermore, we can define an initial probability vector $\pi$ arising from the first application of the channel, with $\sum_{i} \pi_{i} = 1$. The above results imply we can model the protocol classically, by recognizing $T_a$ as $m \times m$ sub-stochastic transition matrices and $\pi$ as an initial stochastic vector (Fig.~\hyperref[fig:protocol]{\ref{fig:protocol}c}). \Cref{eq:probQ} can thus be equivalently written as in \Cref{eq:pC}, with $m$ instead of $d$ states.

We are led to conclude that the repeated inclusion of an arbitrary entanglement-breaking channel before the quantum instrument allows the resulting temporal correlations to be, indeed, described classically. This observation seems to further justify the association of EB channels with classical memories. However, as we already discussed, \emph{any} temporal correlation can be obtained via a classical system if enough memory is available, rendering the previous observation vacuous unless the underlying amount of memory is also taken into account. Thus, any meaningful comparison must be made between systems of \emph{bounded memory}, i.e., bounded dimension.

Here, a key observation is that the effective classical model obtained in \cref{eq:probQ} is over $m$ classical ``virtual'' states (i.e., the number of outcomes in the measure-and-prepare description of the $\cE$), not $d$ as in the dimension of the underlying quantum system. This suggests that a qudit passing through an EB channel may be able to generate stronger correlations than a classical dit. Clearly, the exception is the case $m \le d$, where any quantum model can be converted into a classical one of the same dimension (and vice versa) by choosing $E_i$, $\sigma_i$, and $\cI$ acting diagonally on the same basis; see \cref{app:correspondence} for the explicit construction.

\begin{figure}[t]\centering
	\includegraphics[width=1.0\linewidth]{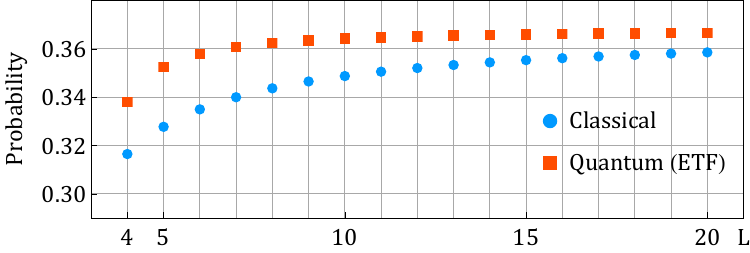}
	\caption{Probabilities of one-tick sequences $\va^L_\text{ot}$ for classical one-way models (cf. Ref.~\cite{budroni2021_tickingclocks,vieira2022_temporal}) and the quantum construction using ETFs, both using $d = L-1$. One-way models are conjectured to be optimal for all $L = d-1$, and are known to achieve the exact upper bounds $\Omega_C(\va^L_\text{ot},d)$ for $L = 3, 4$, and $5$. The bounds for $L = 4$ and $5$ are violated by this explicit quantum construction using ETFs, revealing nonclassical memory effects of EB channels.}
	\label{fig:violations}
\end{figure}

\begin{table}[t]
	\footnotesize
	\begin{tabular}{ |c|c|l|l|l| } 
		\hline
		L & d & \textbf{Classical} ($\Omega_C$) & \textbf{Quantum} ($\Omega_Q$) & 
		$\Omega_Q / \Omega_C$ \\
		\hline
		3 & 2 & $ (2/3)^3 = 0.\overline{296}$ & $\overset{?}{=} 0.296296$ & $\overset{?}{=} 1$ \\
		4 & 3 & $ (3/4)^4 = 0.31640625 $ & $\ge 0.359523$ & $\ge 1.136270$ \\
		5 & 4 & $ (4/5)^5 = 0.32768    $ & $\ge 0.368445$ & $\ge 1.124405$ \\
		\hline
	\end{tabular}
	\caption{Results from numerical optimization for various one-tick sequences of length $L$, and Hilbert space dimension $d = L-1$. The quantum models found numerically for $L = 4$ and $5$ outperform the previous ETF construction, and are included in \cref{app:numopt}. No violations were found for $d = 2$.}
	\label{tbl:violations}
\end{table}

\section{Quantum advantages with entanglement-breaking channels}
Here, we show that EB channels with  $m>d$ are able to generate nonclassical memory effects. This can be investigated by choosing a sequence $\va$ with $d < \DC(\va)$, then computing the maximum probability achievable over all classical models with $d$ states:
\begin{align}
	\Omega_C(\va,d) &= \max_{T} \; p(\va|T,d).
\end{align}
Analogously, using a $d$-dimensional quantum system, we look for instruments $\cI$ and EB channels $\cE$ with $m > d$ such that
\begin{align}\label{eq:pQviol}
	p(\va|\cI,\cE,d) > \Omega_C(\va,d).
\end{align}
Crucially, the maxima $\Omega_C(\va,d)$ must be found explicitly within classical theory, either by exact computation or by obtaining an upper bound, whereas for the quantum value of $p(\va|\cI,\cE,d)$ an explicit model $(\cI,\cE)$ must be found. If  the classical memory bound has been violated even in the presence of an EB channel $\cE$, as in \Cref{eq:pQviol}, then the system still acts as a quantum memory resource for the task of generating temporal correlations.

To achieve this goal, we consider the \emph{one-tick sequences}~\cite{budroni2021_tickingclocks,vieira2022_temporal} of length $L$, defined as $\va_\text{ot}^L := 0\dots01$ and having $\DC(\va^L_\text{ot}) = L$, under the sub-deterministic scenario $d = L-1$. As the case of $L=2$ is trivial, we consider the smallest nontrivial cases $L\geq 3$. For the classical model, analytical upper bounds can be computed via semidefinite programming (SDP) \cite{weilenmann2023_optimisation}. The case $L = 3$ and $4$ were already solved in \cite{weilenmann2023_optimisation}, and here we computed explicitly the case $L=5$ with the same method. In all these cases, the upper bounds computed via SDP coincide, up to the numerical precision, with the values obtained via the \emph{one-way models} in Refs.~\cite{budroni2021_tickingclocks,vieira2022_temporal}, which are conjectured to be optimal for all $d = L-1$ scenarios. We are therefore sure to have obtained the exact bounds for the cases $L=3, 4,$ and $5$. To show EB channels are a quantum memory resource, we must now find an explicit quantum model violating these classical bounds.

One such model can be obtained with the following construction. Consider the deterministic classical model on $m$ states, given by $[T_0]_{i+1,i} = 1$ for $i = 1, \dots, m-1$, with $[T_1]_{1,m} = 1$ and zeros for all other entries, and an initial state $\pi = (1,0,\dots,0)^\T$. The idea is to approximate this model by emulating $m = d+1$ nearly-orthogonal virtual states with the available $d$-dimensional quantum system. With $\zeta = \ee^{2\pi \ii / (m-1)}$ and the basis $\{\ket{k}\}_{k=0}^{d-1}$, let
\begin{equation}
	\ket{\psi_n} = \frac{1}{\sqrt{d-1}} \sum_{k=1}^{d-1} \zeta^{n k} \ket{k}, \quad n = 1, \dots, m-1,
\end{equation}
then define $\cE$ with $\sigma_n = \purestate{\psi_n}$ and $E_{n+1} = \tfrac{d-1}{d} \sigma_n$ for $n = 1, \dots, m-1$, and finally setting $\rho_0 = \sigma_m = E_1 = \purestate{0}$. The instrument is defined as $\cI_a(\rho) = K_a \rho K^\dagger_a$ with $K_0 = \mathrm{diag}(1,\cdots,1,0)$ and $K_1 = \one - K_0$. This quantum model corresponds to isolating the $(d-1)$-dimensional subspace orthogonal to $\ket{0}$, assembling a harmonic complex \emph{equiangular tight frame} (ETF)~\cite{tropp2005} of size $m-1$ within this subspace, then using it for the virtual states in the transitions of $T_0$. ETFs are useful for this task as they provide a generalization of orthonormal bases, saturating the Welch bounds for minimum overlaps~\cite{welch1974}. As seen in \Cref{fig:violations}, this quantum ETF construction clearly violates the exact classical bounds for $L = 4$ and $5$, conclusively showing that general EB channels do not adequately characterize classical memory effects.

The quantum model presented above provides an explicit violation of the classical bound. This is, however, not necessarily the maximum one can achieve in the quantum case. To investigate the quantum maximum with EB channels, using $m = d+1$, we have also performed numerical optimization via gradient descent methods, obtaining lower bounds for the maxima $\Omega_Q(\va,d) = \max_{\cI,\cE} \; p(\va|\cI,\cE,d)$. Results are summarized in Table~\ref{tbl:violations}. The quantum models found numerically slightly outperform the previous ETF construction. More details regarding the optimization and the explicit models can be found in appendices \cref{app:numopt,app:exa}, respectively.

\section{Discussion}
Although EB channels are employed as classical memories in a variety of scenarios, the usefulness of a memory ultimately rests in its ability to generate correlations in time. However, a careful accounting of all memory resources is essential to establish a rigorous notion of nonclassicality for temporal correlations, as the dimension of the memory plays a fundamental role~\cite{zukowski2014_temporal,brierley2015_nonclassicality,clemente2016,budroni2019_memorycost}, and even shared randomness between time steps provides a memory advantage~\cite{frenkel2015_storage,nery2021_simplemaximally,heinosaari2023_simple,taranto2023_hierarchy}. With this in mind, our results show that it is necessary to look beyond qubits and two-time scenarios when investigating memory effects, as otherwise important distinctions between classical and quantum memories are not apparent~\cite{rosset2018_resourcetheory,Simnacher2019PRA,Graffitti2020_PRL,giarmatzi2021_witnessingquantum,yuan2021_benchmarking,ku2022_quantifying}. In other words, memory effects should be recognized as fundamentally multi-time phenomena (as also noted in Refs.~\cite{taranto2023_hierarchy,taranto2025higherorderquantumoperations}), and the dimension of all the memory resources must be explicitly taken into account; see also \cref{app:conditions,app:exa}.

The violations of classical bounds, even in the presence of EB channels, reveal that several established notions of classical and quantum memories are still widely misunderstood, arguably due to improper analogies being drawn from spatial correlations. These analogies ignore the particularities of temporal correlations, such as the necessity of bounding classical memory when establishing a notion of nonclassicality, leading to an inadequate treatment of the relevant memory resources. Concretely, while the dimension $m$ of the intermediate classical space of an EB channel plays a fundamental role in its original temporal context---namely, bounding the amount of classical communication between inputs and outputs---, its role is obscured in typical applications of EB channels, where the value of $m$ does not affect the separability of the resulting quantum supermap.

Indeed, in light of our findings, even if unlimited classical communication is considered a free resource, EB channels still cannot be considered ``classical memories''. Therefore, the common presumption that EB channels generically act as a classical resource carries additional assumptions, which could be rather strong; see discussion on \cref{app:conditions}. More generally, EB channels should be understood as quantum processes with classical memory~\cite{taranto2025higherorderquantumoperations}, while also highlighting their interplay between quantum and classical resources. Our results show this interplay leads to subtle nonclassical memory effects, a phenomenon which warrants further attention.

\section{Conclusions}
Entanglement-breaking channels are universally recognized as classical resources in a variety of contexts, ranging from quantum networks~\cite{rosset2018_resourcetheory,Simnacher2019PRA,Graffitti2020_PRL,yuan2021_benchmarking,giarmatzi2021_witnessingquantum,ku2022_quantifying,Ku2023arXiv,Abiuso_2024}, to multi-time quantum processes and open quantum systems~\cite{milz2021_genuine,berk2021_resourcetheoriesof,roy2024_semidevice,santos2024_quantifying,berk2021_resourcetheoriesof,taranto2023_hierarchy,backer2024_localdisclosure}. Quite surprisingly, however, we have found sequential scenarios where they still allow for genuinely quantum memory effects, leading to nonclassical temporal correlations. More precisely, we answer in the negative the following question: is a qudit passing through an entanglement-breaking channel equivalent to a classical dit?

To address this question, we have investigated a task where memory is the central resource, namely, the generation of correlations in time. Indeed, it has been shown that any temporal correlation can be reproduced by a classical system if enough memory is available~\cite{fritz2010_temporalCHSH,clemente2016,hoffmann2018_temporalqubit, spee2020_simulating}, such that differences between classical and genuinely quantum temporal correlations only emerge if memory is bounded~\cite{zukowski2014_temporal,brierley2015_nonclassicality,budroni2019_memorycost,budroni2021_tickingclocks,vieira2022_temporal}---a fundamental but often neglected feature of temporal correlations. As classical memory is understood in terms of the number of states $d$, a fair comparison between classical and quantum theory must involve a quantum system of the same size. Differences between quantum and classical memories can be found by investigating the most general set of memory operations allowed in each theory, then establishing bounds on their capabilities.

Our findings inspire several future research directions. Understanding the minimal requirements for an EB channel to truly act as a classical memory could offer further insights into the origins of quantum advantages, allowing a more definitive understanding of classical and quantum memory effects; see also \cref{app:correspondence}. Furthermore, since the quantum advantage in our case emerges from the repeated interplay between states $\sigma_i$ and measurements $E_i$, an investigation of the quantum upper bounds in the presence of arbitrary EB channels may inspire new strategies for optimal encoding of classical information within quantum systems in various sequential tasks, with potential applications to, e.g., quantum random-access codes~\cite{wiesner1983,ambainis1999,tavakoli2015qracs,miklin2020,roy2024_semidevice}, or the quantum compression of classical predictive models for stochastic processes~\cite{binder2018,elliott2022,wu2023}. 
In addition, even if in this work we have focused on sequential measurements on a single system, our approach for investigating classical and quantum resources via finite-state machines can be naturally extended to arbitrary spatio-temporal scenarios. Indeed, a related result has been recently found in entanglement-assisted communication scenario \cite{ChiribellaPRL2025}, namely, that the communication of classical messages through a noisy entanglement-breaking qubit channel assisted by quantum entanglement cannot, in general, be simulated by communication through a noisy bit channel assisted by classical correlations. By enabling an exact characterization of memory and communication resources needed for any given task, this approach can lead to novel insights even in simple scenarios, as was the case here.

Naturally, if a quantum memory is required to preserve entanglement, then EB channels are not a useful resource. However, quantum effects and their advantages are not restricted to entanglement~\cite{dakic2014_discord,pirandola2014_discord} or spatial correlations~\cite{zukowski2014_temporal,budroni2019_memorycost,mansfield2018_contextuality,howard2014,saha2023_incompatibility}. By recognizing the temporal character inherent to any information processing task, we are required to also consider the fundamental differences between classical and quantum temporal correlations.
Our results indicate a more careful and detailed investigation of quantum memory resources is warranted, such that their precise origin and advantage over classical memories can be properly understood, quantified, and successfully applied in emergent quantum technologies.

\section*{Acknowledgements}
We thank Simon Milz, Philip Taranto, Marco Túlio Quintino and Felix Binder for valuable discussions. This work is supported by the Austrian Science Fund (FWF) through projects ZK 3 (Zukunftskolleg) and F7113 (BeyondC), and partially funded within the QuantERA II Programme that has received funding from the EU’s H2020 research and innovation programme under the GA No 101017733. H.-Y. K. is supported by the National Science and Technology Council (NSTC), (with grant number NSTC 112-2112-M-003- 020-MY3), and the MOE through Higher Education Sprout Project of National Taiwan Normal University.

\appendix
\crefalias{section}{appendix}
\section{Direct correspondence between classical and quantum models for $m \le d$}\label{app:correspondence}

Using $\cA$ for the set of outcomes, let $T = (T_a)_{a \in \cA}$ and $\{\ket{i}\}_{i=1}^d$ denote an arbitrary classical model and orthonormal basis, respectively. We consider a generic quantum instrument written in its Kraus decomposition
\begin{equation}
	\cI_a(\rho) = \sum_{k=1}^{n_a} K_{ak} \rho K_{ak}^\dagger, \quad \text{with}\quad \sum_{a \in \cA} \sum_{k=1}^{n_a} K_{ak}^\dagger K_{ak} = \one,
\end{equation}
such that $\sum_a \cI_a$ is trace preserving. We can define a Kraus operator for each non-zero $[T_a]_{ji}$, via the action $K_{aij} \ket{i} = \sqrt{[T_a]_{ji}} \ket{j}$, such that $\cI$ acts diagonally in this basis. Now, by choosing $E_i = \sigma_i = \purestate{i}$, we have a completely dephasing channel $\cE_\text{CD}$ in the same basis, $\cE_\text{CD}(\rho) := \sum_{i=1}^m \bra{i} \rho \ket{i} \purestate{i}$. Finally, by using the initial state $\rho_0 = \sum_i \pi_i \purestate{i}$, we conclude
\begin{equation}\label{eq:pCeqpQ}
	\Tr{ \cI_{a_L} \circ \cE \circ \cdots \circ \cI_{a_1} \circ \cE(\rho_0) } = \eta T_{a_L} \cdots T_{a_1} \pi.
\end{equation}
Together with the results from \Cref{eq:probQ,eq:ItoT}, this establishes the correspondence between classical and quantum scenarios for $m \le d$, and motivates the investigation of the $m > d$ case.

\section{Conditions for classicality}\label{app:conditions} Since generic EB channels can violate bounds for classical memories, it is desirable to know the conditions under which nonclassical memory effects are actually erased by their application, as often assumed, i.e., such that \emph{no} choice of quantum instrument $\cI$ can lead to nonclassical temporal correlations.

A full investigation of these conditions is beyond the scope of this paper. Nevertheless, understanding some sufficient conditions offers insights on the strength of the implicit assumptions being made when applying EB channels as means of characterizing classical memory effects. Assuming input and output spaces have the same number of orthogonal states, we present two independent sufficient conditions: either the states $\{\sigma_i\}_{i=1}^m$ commute, and/or the POVM elements $\{E_i\}_{i=1}^m$ commute.

If $[\sigma_i, \sigma_j] = 0$, $\forall i,j$, then a common eigenbasis exists in which all operators can be written as $\sigma_i= \mathrm{diag}(s^0_i,\dots,s^{d-1}_i)$. Without loss of generality, we can express the channel in this basis as
\begin{equation}
	\begin{split}
		\cE(\rho) = \sum_{i=1}^{m} \Tr{E_i \rho} \sigma_i = \sum_{i=1}^m \sum_{\ell=0}^{d-1} \Tr{E_i \rho } s_i^\ell \purestate{\ell}\\
		= \sum_{\ell=0}^{d-1} \Tr{ \left(\sum_{i=1}^m s_i^\ell E_i\right) \rho }  \purestate{\ell}= \sum_{\ell=0}^{d-1} \Tr{ F_\ell \rho }  \purestate{\ell},
	\end{split}
\end{equation}
where $F_\ell:= \sum_i s_i^\ell E_i$ and $\{F_\ell\}_{\ell=1}^d$ is a valid POVM.  In fact, $s_i^\ell \geq 0$  since $\sigma_i\geq 0$ and $\sum_\ell s_i^\ell=1$ since $\Tr{\sigma_i}=1$,  which implies $F_\ell \geq 0$ and $\sum_\ell F_\ell=\sum_{i,\ell}s_i^\ell E_i= \sum_{i} E_i= \openone$. We thus have a new definition of the channel $\cE$ which uses only $d$ outcomes and states. But from \cref{app:correspondence}, this implies there exists an equivalent classical model with $d$ states. Similarly, if $[E_i, E_j] = 0$, $\forall i,j$, we may write $E_i = \sum_{\ell=0}^{d-1} e_i^\ell \purestate{\ell}$, such that
\begin{equation}
	\begin{split}
		\cE(\rho) = \sum_{i=1}^{m} \Tr{E_i \rho} \sigma_i = \sum_{i=1}^m \sum_{\ell=0}^{d-1} e_i^\ell \Tr{\rho \purestate{\ell}} \sigma_i \\
		= \sum_{\ell=0}^{d-1} \Tr{\rho \purestate{\ell}} \left( \sum_{i=1}^{m} e_i^\ell \sigma_i \right) = \sum_{\ell=0}^{d-1} \Tr{\rho \purestate{\ell}} \tilde{\sigma}_\ell,
	\end{split}
\end{equation}
where $\tilde{\sigma}_\ell := \sum_{i=1}^{m} e_i^\ell \sigma_i$ are valid states, since $e_i^\ell \ge 0$ and $\sum_{i=1}^m e_i^\ell = 1$ for every $\ell$, and we once again obtain a channel $\cE$ written in terms of $d$ outcomes.

The above results hold under the assumption that input and output spaces of $\cE$ have the same dimension. To understand why this is necessary, consider two separate EB channels, the quantum-classical $\cE_1(\rho) = \sum_{i=1}^{m} \Tr{\rho E_i} \ketbra{i}{i}$ and the classical-quantum $\cE_2(\rho) = \sum_{i=1}^{m} \Tr{\ketbra{i}{i} \rho} \sigma_i$, such that $\cE = \cE_2 \circ \cE_1$. The output states of $\cE_1$ and POVM elements of $\cE_2$ are both $\{\ketbra{i}{i}\}_{i=1}^m$, which are completely classical and within an $m$-dimensional space. Thus, taken individually, it would seem that both $\cE_1$ and $\cE_2$ should be completely classical. Nevertheless, their concatenation leads to nonclassical temporal correlations, as we have shown. The fact $m > d$ has allowed subtle quantum effects to emerge from the interplay between sequential encoding and decoding of quantum information into the classical memory. The sequential nature of our scenario prevents a straightforward analysis, however, as nonclassical effects only emerge beyond qubits and at least four time steps.

We also note that these conditions for classicality are reminiscent to those for zero quantum discord in bipartite separable states~\cite{dakic2010_discord}, suggesting that non-zero discord of the Choi state of an EB channel may play a role in its temporal nonclassicality.  We leave such investigations to future research.

\section{Numerical optimization}\label{app:numopt}

For $\va^L_\text{ot} = 0\dots01$ and $d = L-1$, the best classical bounds known are achieved by the so-called one-way models~\cite{budroni2019_memorycost,vieira2022_temporal}
\begin{equation}
	[T_0]_{i,i} = 1/L, \quad [T_0]_{i,i+1} =  [T_1]_{d,1} = 1 - 1/L,
\end{equation}
with $p(\va^L_\text{ot}|T,d) = (1 - \sfrac{1}{L})^L$, which are conjectured to give the optimal probability for a given $d$ in any sub-deterministic classical scenario (i.e., $d < \DC(\va)$). For $d = 2,3$, and $4$, these upper bounds are known to be exact, and can be computed with semidefinite programming techniques~\cite{weilenmann2023_optimisation}. 

In the quantum case, numerical optimization was performed via gradient descent techniques using the Adam algorithm~\cite{kingma2017_adam}, with an additional exponentially decreasing learning rate from $0.07$ to $10^{-12}$ over 50\,000 iterations, and many trials with random initial values. We have found sufficient to optimize for rank-1 $E_i$ and $ \sigma_i$, with no further assumptions. Each $\cI_a$ was defined with a single Kraus operator, as no advantage was observed in using more. As the probability in \Cref{eq:probQ} is convex for $\rho_0$, and we maximize for a single sequence, this initial state can be assumed pure, and w.l.o.g., set to $\rho_0 = \purestate{0}$. We have  the following constrained optimization problem:
\begin{align}\label{eq:optproblem}\begin{split}
		\textbf{Find:} & \quad \max_{\cE,\cI} p(\va|\cE,\cI,d) \\
		\textbf{Such that:} & \quad E_i, \sigma_i \ge 0 \; \forall i, \\
		& \quad \sum_i E_i = \one, \; \Tr{\sigma_i} = 1, \; \forall i, \\
		& \quad \sum_{a} K_a^\dagger K_a = \one,
\end{split}\end{align}
with $\cE(\rho) = \sum_{i=1}^m \Tr{\rho E_i} \sigma_i$ and $\cI_a(\rho) = K_a \rho K_a^\dagger$. This can be relaxed into an unconstrained optimization by using arbitrary $d \times d$ complex matrices $A_i, B_i, C_a$, then defining $\sigma_i := A_i^\dagger A_i / \mathrm{tr}[A_i^\dagger A_i]$ for the states. For the POVM, we first define $\tilde{B}_i := B_i^\dagger B_i$ and compute $\lambda^E_\text{max} := \max \text{eigs}(\sum_i \tilde{B}_i)$, such that letting $E_i := \tilde{B}_i / \lambda^E_\text{max}$ gives $\sum_i E_i \le \one$. For the instrument we compute $\lambda^\cI_\text{max} =  \max \text{eigs}( \sum_a C_a^\dagger C_a )$ and define $K_a := C_a / \sqrt{\lambda^\cI_\text{max}}$, giving $\sum_a K_a^\dagger K_a \le \one$. In this way, the gradient of the objective function will naturally favor solutions satisfying $\sum_i E_i = \one$ and $\sum_a K_a^\dagger K_a = \one$. The above construction is general, but can be easily adapted for rank-1 $\sigma_i$ and $E_i$. The best quantum models found for $L = 4$ and $5$ are included in \cref{app:exa}, below.

\section{Numerical examples violating the classical bound}\label{app:exa}
The following are the best quantum models found for $L = 4$ and $5$ using gradient descent techniques, as described in \cref{app:numopt}. These optimal values for a given dimension and sequence length required high accuracy numerical optimization to be found, and rely on several significant figures in order to adequately satisfy the constraints in \Cref{eq:optproblem}. Despite considerable efforts, no closed-form description of these models was found. Since all optimal $E_i$ and $\sigma_i$ found were rank-1, we include the vectors such that $E_i = \purestate{e_i}$ and $\sigma_i = \purestate{\phi_i}$. We provide only 5 significant digits of precision, for simplicity, but convergence was obtained up to 10 significant digits. Both cases assume $\rho_0 = \purestate{0}$ and $\cI_a(\rho) = K_a \rho K_a^\dagger$. For $L = 3$ and $d = 2$, we were unable to find any violation of the classical bound, strongly suggesting EB-based temporal nonclassicality only emerges beyond two-level systems. Furthermore, since $\DC(\va) \le L$ for any sequence of length $L$, either a qubit or a bit are sufficient to produce any correlations on two time steps. It is therefore essential to also look beyond two-time steps in order to reveal fundamental differences between classical and quantum systems in this sequential measurement scenario.

\onecolumngrid
\textbf{Model for $L = 4$}:
\begin{align}\begin{split}
&\ket{e_1} = \begin{bmatrix} 1 \\ 0 \\ 0 \end{bmatrix},\quad \ket{e_2} = \begin{bmatrix} 0 \\ -0.09692-0.41924 i \\ 0.74404 \end{bmatrix}, \quad \ket{e_3} = \begin{bmatrix} 0 \\ 0.07184\, +0.30898 i \\ 0.65475 \end{bmatrix}, \quad \ket{e_4} = \begin{bmatrix} 0 \\ -0.18847-0.82383 i \\ -0.13307 \end{bmatrix}, \\
&\ket{\phi_1} = \begin{bmatrix} 0 \\ -0.08293-0.35949 i \\ 0.92946 \end{bmatrix}, \quad \ket{\phi_2} = \begin{bmatrix} 0 \\ 0.09971\, +0.4288 i \\ 0.89788 \end{bmatrix}, \quad \ket{\phi_3} = \begin{bmatrix} 0 \\ 0.96462 \\ 0.05913\, -0.25695 i \end{bmatrix}, \quad \ket{\phi_4} = \
\begin{bmatrix} 1 \\ 0 \\ 0 \end{bmatrix}, \\
&K_0 = \mathrm{diag}(0,1,1), \qquad K_1 = \mathrm{diag}(1,0,0).
\end{split}\end{align}

\textbf{Model for $L = 5$}:
\begin{align}\begin{split}
&\ket{e_1} = \begin{bmatrix} 1 \\ 0 \\ 0 \\ 0 \end{bmatrix} ,\quad \ket{e_2} = \begin{bmatrix} 0 \\ \
	-0.18168-0.03287 i \\ -0.33582-0.01918 i \\ 0.83859 \end{bmatrix} ,\quad \ket{e_3} = \
\begin{bmatrix} 0 \\ 0.06342\, +0.62645 i \\ -0.36075+0.0699 i \\ -0.32475 \end{bmatrix} ,\quad \
\ket{e_4} = \begin{bmatrix} 0 \\ -0.40607-0.42989 i \\ -0.28345-0.24945 i \\ -0.4251 \end{bmatrix} ,\\
&\ket{e_5} = \begin{bmatrix} 0 \\ -0.3906+0.2592 i \\ 0.78055 \\ 0.05633\, +0.08616 i \
\end{bmatrix}, \quad
\ket{\phi_1} = \begin{bmatrix} 0 \\ -0.25627-0.10917 i \\ -0.29951-0.06935 i \\ 0.90988 \
\end{bmatrix} ,\quad \ket{\phi_2} = \begin{bmatrix} 0 \\ 0.78968 \\ 0.17386\, +0.37575 i \\ \
	-0.02481+0.45209 i \end{bmatrix} ,\\
&\quad \ket{\phi_3} = \begin{bmatrix} 0 \\ 0.73235 \\ 0.46844\, \
	-0.13164 i \\ 0.35015\, -0.32293 i \end{bmatrix} ,\quad \ket{\phi_4} = \begin{bmatrix} 0 \\ \
	-0.34778+0.40375 i \\ 0.82555 \\ 0.14552\, +0.11544 i \end{bmatrix} ,\quad \ket{\phi_5} = \
\begin{bmatrix} 1 \\ 0 \\ 0 \\ 0 \end{bmatrix}, \\
&K_0 = \mathrm{diag}(0,1,1,1), \qquad K_1 = \mathrm{diag}(1,0,0,0).
\end{split}\end{align}

\bibliography{references}{}

\end{document}